\documentclass[aps,prc, nofootinbib,twocolumn]{revtex4-1}

\usepackage{epsfig}
\usepackage{color}
\usepackage[latin1]{inputenc}
\usepackage{float,amsmath}
\usepackage{graphicx}

\newcommand{\be}{\begin{eqnarray}}
\newcommand{\ee}{\end{eqnarray}}

\newcommand{\benum}{\begin{enumerate}}
\newcommand{\eenum}{\end{enumerate}}

\begin{document}

\title{Higher flow harmonics from (3+1)D event-by-event viscous hydrodynamics}

\author{Bj\"orn Schenke}
\affiliation{Physics Department,  Brookhaven National Laboratory,  Upton, NY-11973, USA}

\author{Sangyong Jeon}
\affiliation{Department of Physics, McGill University, 3600 University Street, Montreal, Quebec, H3A\,2T8, Canada}

\author{Charles Gale}
\affiliation{Department of Physics, McGill University, 3600 University Street, Montreal, Quebec, H3A\,2T8, Canada}

\begin{abstract}
We present event-by-event viscous hydrodynamic calculations of the anisotropic flow coefficients $v_2$ to $v_5$ for heavy-ion collisions at the Relativistic Heavy-Ion Collider (RHIC).
We study the dependence of different flow harmonics on shear viscosity and the morphology of the initial state. $v_3$ and higher flow harmonics exhibit a particularly strong dependence on both the initial granularity and shear viscosity. We argue that a combined analysis of all available flow harmonics will allow to determine $\eta/s$ of the quark gluon plasma precisely. Presented results strongly hint at a value $(\eta/s)_{QGP}<2/4\pi$ at RHIC.
Furthermore, we demonstrate the effect of shear viscosity on pseudo-rapidity spectra and the mean transverse momentum as a function of rapidity. 
\end{abstract}

\maketitle


\section{Introduction}
Hydrodynamics is an indispensable and accurate tool for the description of the
bulk behavior of a fluid.
The equations of hydrodynamics are just the
conservation laws, an additional equation of state and
constitutive relationships for dissipative hydrodynamics.
The idea that ideal hydrodynamics can describe the 
outcome of hadronic collisions has a long history.
Applications to relativistic heavy-ion collisions
have been carried out by many researchers (see \cite{Schenke:2010nt, Schenke:2011qd} for an extensive list of references).

Fluctuating initial conditions for hydrodynamic simulations of heavy-ion collisions
have been argued to be very important for the exact determination of collective flow observables and to describe 
features of multi-particle correlation measurements in heavy-ion collisions
\cite{Andrade:2006yh,Adare:2008cqb,Abelev:2008nda,Alver:2008gk,Alver:2009id,Abelev:2009qa,Miller:2003kd,Broniowski:2007ft,Andrade:2008xh,Hirano:2009bd,
Takahashi:2009na,Andrade:2009em,Alver:2010gr,Werner:2010aa,Holopainen:2010gz,Alver:2010dn,Petersen:2010cw,Schenke:2010rr,Schenke:2011tv,Qiu:2011iv}.
Real event-by-event hydrodynamic simulations have been performed and show modifications to spectra and flow from ``single-shot'' 
hydrodynamics with averaged initial conditions \cite{Holopainen:2010gz,Schenke:2010rr,Schenke:2011tv,Qiu:2011iv}.
An important advantage of event-by-event hydrodynamic calculations is the possibility to consistently study all higher flow harmonics
in the same simulation without the need for an artificial construction of an initial eccentricity, triangularity, etc.
This is particularly important for the computation of $v_4$, which receives strong contributions from elliptical deformations of the initial state,
and $v_5$, which couples to triangularity from fluctuations and to the ellipticity of the collision geometry \cite{Qiu:2011iv}. Recent 3+1D viscous hydrodynamic simulations have highlighted the role of fluctuating initial states also on electromagnetic observables \cite{Dion:2011pp}. 

Different $v_n$ depend differently on $\eta/s$ and the details of the initial condition, 
which is determined by the dynamics and fluctuations of partons in the incoming nuclear wave functions.
In this work we present quantitative results on the dependence of $v_2$ to $v_5$ on both the shear viscosity to entropy density ratio $\eta/s$
and the granularity of the initial state, and compare to experimental data.

This paper is organized as follows. In Section \ref{vischyd} we introduce the employed second order relativistic viscous hydrodynamic framework. 
The explicit form of the hyperbolic equations in $\tau$-$\eta_s$ coordinates and the numerical implementation are presented in Section \ref{impl}.
We discuss the initial condition for single events in Section \ref{ini} and explain the freeze-out procedure in Section \ref{freeze}.
Finally, results are presented in Section \ref{results}, followed by conclusions and discussions in Section \ref{summary}.


\section{Viscous hydrodynamics}\label{vischyd}
In \cite{Schenke:2010nt} we introduced the simulation \textsc{music} for ideal relativistic fluids and extended
it in \cite{Schenke:2010rr} to include dissipative effects.

In the ideal case, the evolution of the system, created in relativistic heavy-ion collisions,
is described by the following 5 conservation equations
\begin{eqnarray}
& \partial_\mu T_{\rm id}^{\mu\nu} = 0\,,\label{conservationEqns1} 
\\
& \partial_\mu J_B^\mu = 0\,, \label{conservationEqns2}
\end{eqnarray}
where $T_{\rm id}^{\mu\nu}$ is the energy-momentum tensor and $J_B^\mu$ is the net baryon
current. These are usually re-expressed
using the time-like flow 4-vector $u^\mu$ as
\begin{eqnarray}\label{Tideal}
& T_{\rm id}^{\mu\nu} = (\varepsilon + \mathcal{P})u^\mu u^\nu - \mathcal{P} g^{\mu\nu}\,,
\\
& J_{B}^{\mu} = \rho_{B} u^\mu \,,
\end{eqnarray}
where $\varepsilon$ is the energy density, $\mathcal{P}$ is the pressure, $\rho_B$
is the baryon density and 
$g^{\mu\nu} = \hbox{diag}(1, -1, -1, -1)$ is the metric tensor.
The equations are then closed by adding the equilibrium 
equation of state
\begin{eqnarray}
\mathcal{P} = \mathcal{P}(\varepsilon, \rho_B)
\end{eqnarray}
as a local constraint on the variables.

Historically, these equations have first been solved in a boost-invariant framework \cite{Bjorken:1982qr},
eliminating the longitudinal direction and assuming uniformity in the transverse direction.
At RHIC the central plateau in rapidity extends over 4 units.
Hence, as long as one is concerned only with the dynamics near the mid-rapidity region, boost invariance should be a valid approximation at RHIC, 
restricting the relevant spatial dimensions to the transverse plane. 
Much success has been achieved by these (2+1)D calculations (see references in \cite{Schenke:2010nt} and \cite{Huovinen:2003fa,Kolb:2003dz} for thorough
reviews).
However, in order to analyze experimental data away from mid-rapidity, inclusion of the non-trivial longitudinal dynamics is essential \cite{Aguiar:2000hw,Hirano:2001yi,Hirano:2001eu,Hirano:2002ds,Nonaka:2000ek,Nonaka:2006yn,Schenke:2010nt}.


The next step in improving relativistic hydrodynamic simulations of heavy-ion collisions is the inclusion of finite viscosities.
In the first order, or Navier-Stokes formalism for viscous hydrodynamics,
the stress-energy tensor is decomposed into
\be
T_{\rm 1st}^{\mu\nu} = T^{\mu\nu}_{\rm id} + S^{\mu\nu}\,,
\ee
where
$T_{\rm id}^{\mu\nu}$ is given by Eq.\,(\ref{Tideal})


The viscous part of the stress energy tensor in the first-order approach is
given by 
\be
S^{\mu\nu} = \eta 
\left(
\nabla^\mu u^\nu + \nabla^\nu u^\mu - 
{2\over 3}\Delta^{\mu\nu}\nabla_\alpha u^\alpha
\right)
\ee
where $\Delta^{\mu\nu} = g^{\mu\nu} - u^\mu u^\nu$ is the local 3-metric 
and $\nabla^\mu = \Delta^{\mu\nu}\partial_\nu$ is the local spatial derivative.
Note that $S^{\mu\nu}$ is transverse with respect to the flow velocity since
$\Delta^{\mu\nu}u_\nu = 0$ and $u^\nu u_\nu = 1$. 
Hence, $u^\mu$ is also an
eigenvector of the whole stress-energy tensor with the same eigenvalue
$\epsilon$. $\eta$ is the shear viscosity of the medium, which we assume to be constant.

This form of viscous hydrodynamics is conceptually simple. 
However, this Navier-Stokes form is known to introduce unphysical 
super-luminal signals \cite{Hiscock:1983zz,Hiscock:1985zz,Pu:2009fj}, leading to numerical instabilities. 
The second-order Israel-Stewart formalism \cite{Israel:1976tn,Stewart:1977,Israel:1979wp}
avoids this super-luminal propagation, as does the more recent approach in 
\cite{Muronga:2001zk}. 

In this work, we use a variant of the Israel-Stewart formalism
derived in \cite{Baier:2007ix}, where the stress-energy tensor is decomposed as
\be\label{tmunu}
{\cal T}^{\mu\nu} = T^{\mu\nu}_{\rm id} + \pi^{\mu\nu}\,.
\ee
The evolution equations are
\begin{equation}
\partial_\mu {\cal T}^{\mu\nu} = 0
\label{eq:idealEq}
\end{equation}
and
\begin{equation}
\Delta^{\mu}_{\alpha}\Delta^{\nu}_{\beta}
{u^\sigma\partial_\sigma} \pi^{\alpha\beta}
=
-{1\over \tau_\pi}
\left( \pi^{\mu\nu} - S^{\mu\nu} \right) - {4\over 3}\pi^{\mu\nu}(\partial_\alpha u^\alpha)\,.
\label{eq:Weq}
\end{equation}


When dealing with rapid longitudinal expansion,
it is useful to transform these equations to the $\tau$-$\eta_s$-coordinate system, defined by
\begin{eqnarray}
t &=& \tau\cosh\eta_s\,,\nonumber\\
z &=& \tau\sinh\eta_s\,.
\label{eq:tz}
\end{eqnarray}

We obtain the following hyperbolic equations with sources
\be
\partial_a T_{\rm id}^{ab} = -\partial_a \pi^{ab} + F^b
\label{eq:Tid_eq}
\ee
and
\be
\partial_a (u^a \pi^{cd}) = -(1/\tau_\pi)(\pi^{cd} - S^{cd}) + G^{cd}
\label{eq:uW_eq}
\ee
where $F^b$ and $G^{cd}$ contain terms introduced by the coordinate change
from $t,z$ to $\tau,\eta_s$ as well as those introduced by the projections in
Eq.\,(\ref{eq:Weq}), and $\tau_\pi$ is the relaxation time.

Our approach to solve these hyperbolic equations relies on
the Kurganov-Tadmor (KT) scheme
\cite{Kurganov:2000,Naidoo:2004}, together
with Heun's method to solve resulting ordinary
differential equations.





\section{Implementation}\label{impl}
 As mentioned above, the most natural coordinate system for us is the $\tau-\eta_s$ coordinate
 system defined by Eq.\,(\ref{eq:tz}).
 In this coordinate system, the conservation equation $\partial_\mu J^\mu = 0$ becomes
 \be
  \partial_\tau (\tau J^\tau)
 + \partial_v(\tau J^v) 
+\partial_{\eta_s} J^{\eta_s}& = &0\,,
 \label{eq:jcons}
 \ee
 where
 \be
 J^\tau &= & (\cosh\eta_s J^0 - \sinh\eta_s J^3)\,,
 \label{eq:jtau}
 \\
 J^{\eta_s} &= &
 ( \cosh\eta_s J^3 - \sinh\eta_s J^0 )\,,
 \label{eq:jeta}
 \ee
 which is simply a Lorentz boost with the space-time rapidity
 $\eta_s = \tanh^{-1}(z/t)$. 
 The index $v$ and $w$ in this
 section always refer to the transverse $x,y$ coordinates which are not
 affected by the boost.
 Applying the same transformation to both indices in Eq.\,(\ref{eq:idealEq}),
 one obtains
\begin{align}\label{eq:econs}
& \partial_\tau (\tau T^{\tau\tau})
+\partial_v (\tau T^{v\tau}) +\partial_{\eta_s} (T^{\eta_s\tau}) 
+T^{\eta_s\eta_s}\\
&+\partial_\tau (\tau \pi^{\tau\tau}) +\partial_v (\tau \pi^{v\tau}) +\partial_{\eta_s} (\pi^{\eta_s\tau}) 
+ \pi^{\eta_s\eta_s} = 0\,,\nonumber
\end{align}
\begin{align}\label{eq:etacons}
&\partial_\tau (\tau T^{\tau\eta_s})
+ \partial_v(\tau T^{v\eta_s}) + \partial_{\eta_s}(T^{\eta_s\eta_s})
+ T^{\tau\eta_s}\\
&+ \partial_\tau (\tau \pi^{\tau\eta_s})
+ \partial_v(\tau \pi^{v\eta_s}) + \partial_{\eta_s}(\pi^{\eta_s\eta_s})
+ \pi^{\tau\eta_s} = 0\,,\nonumber
\end{align}
and
\begin{align}\label{eq:xycons}
&\partial_\tau (\tau T^{\tau v})
+
\partial_w (\tau T^{w v})
+ 
\partial_{\eta_s} (T^{\eta_s v})\\
&+ \partial_\tau (\tau \pi^{\tau v})
+
\partial_w (\tau \pi^{w v})
+ 
\partial_{\eta_s} (\pi^{\eta_s v}) = 0
\,.\nonumber
\end{align}
These 5 equations, namely Eq.\,(\ref{eq:jcons}) for the net baryon current, and
Eqs.\,(\ref{eq:econs}, \ref{eq:etacons}, \ref{eq:xycons}) for the energy and
momentum, are solved along with Eqs.\,(\ref{eq:uW_eq}) for the viscous part of the stress-energy tensor, which in a more explicit way of writing read

\begin{align}\label{eq:uW_eqExplicit}
  \partial_c(u^c \pi^{ab}) = &-\frac{1}{2 \tau}u^\tau \pi^{ab} + \frac{1}{\tau}\Delta^{a\eta}u^\eta \pi^{b \tau} - \frac{1}{\tau} \Delta^{a\tau} u^\eta \pi^{b\eta} 
  \nonumber\\
  & - g_{cf} \pi^{cb}u^a D u^f - \frac{\pi^{ab}}{2 \tau_\pi} - \frac{1}{6}\pi^{ab}\partial_c u^c\nonumber\\
  & + \frac{\eta}{\tau_\pi}\left(-\frac{1}{\tau}\Delta^{a\eta}g^{b\eta}u^\tau + \frac{1}{\tau}\Delta^{a\eta}g^{b\tau}u^\tau \right.\nonumber\\
  & ~~~~~~~~~~ \left. + g^{ac}\partial_c u^b - u^a D u^b - \frac{1}{3} \Delta^{ab}\partial_c u^c \right)\nonumber\\
  & + (a\leftrightarrow b)\,,
\end{align}
The relaxation time $\tau_\pi$ is set to $3\eta/(\epsilon+\mathcal{P})$, in line with the approach in \cite{Song:2007ux}. 
It was also shown in \cite{Song:2009gc} that the dependence of observables such as $v_2$ on $\tau_\pi$ is negligible when including the
term $(4/3)\pi^{\mu\nu}(\partial_\alpha u^\alpha)$ in Eq.\,(\ref{eq:uW_eq}).

To solve the equations we use the KT algorithm as explained in \cite{Schenke:2010nt}.
In detail, we compute the first step within Heun's method for Eqs.\,(\ref{eq:jcons}, \ref{eq:econs}, \ref{eq:etacons}, \ref{eq:xycons}),
then the first step for Eqs.\,(\ref{eq:uW_eqExplicit}), proceed with the second step for
Eqs.\,(\ref{eq:jcons}, \ref{eq:econs}, \ref{eq:etacons}, \ref{eq:xycons}) using the evolved result for $\pi^{ab}$, and finally compute the
second step for Eqs.\,(\ref{eq:uW_eqExplicit}). This concludes the evolution of one time step.

One major difference to the ideal hydrodynamic equations solved in \cite{Schenke:2010nt} is the appearance of
time derivatives in the source terms of Eqs.\,(\ref{eq:econs}, \ref{eq:etacons}, \ref{eq:xycons}, \ref{eq:uW_eqExplicit}). 
These are handled with the first order approximation
\begin{equation}\label{eq:timederiv1}
\dot{g}(\tau_n) = (g(\tau_{n})-g(\tau_{n-1}))/\Delta \tau\,,
\end{equation}
in the first step of the Heun method,
and in the second step we use
\begin{equation}\label{eq:timederiv2}
\dot{g}(\tau_n) = (g^*(\tau_{n+1})-g(\tau_n))/\Delta \tau\,,
\end{equation}
where $g^*(\tau_{n+1})$ is the result from the first step.

As in most Eulerian algorithms, ours also suffers from numerical
instability when the density becomes small while the flow velocity becomes
large. Fortunately this happens late in the evolution or at the very edge of the system.
Regularizing such instability has no strong effects on the observables we are interested in. 
Some ways of handling this are known (for instance see Ref.\cite{Duez:2004nf}). 

In this study, when finite viscosity causes negative pressure in the cell, we revert
to the previous value of $\pi^{\mu\nu}$ and reduce all components by 5\%.
This procedure stabilizes the calculations without introducing spurious effects.

\section{Initialization and Equation of State}\label{ini}


To determine the initial energy density distribution for a single event, we employ the Monte-Carlo Glauber model using the method
described in \cite{Alver:2008aq} to determine the initial distribution of wounded nucleons. 
Before the collision the density distribution of the two nuclei is described by
a Woods-Saxon parametrization, which we sample to determine the positions of individual nucleons.
The impact parameter is sampled from the distribution 
\begin{equation}
P(b) db = 2bdb/(b_{\rm max}^2-b_{\rm min}^2)\,,
\end{equation}
where $b_{\rm min}$ and $b_{\rm max}$ depend on the given centrality class.
Given the sampled initial impact parameter the two nuclei are superimposed.
Two nucleons are assumed to collide if their relative transverse distance is less than
\begin{equation}
 D = \sqrt{\sigma_{NN}/\pi}\,,
\end{equation}
where $\sigma_{NN}$ is the inelastic nucleon-nucleon cross-section, which at top RHIC energy of $\sqrt{s}=200A\,{\rm GeV}$ is $\sigma_{NN}=42\,{\rm mb}$.
The energy density is taken to scale mostly with the wounded nucleon distribution and to 25\% with the binary collision distribution. 
So, two distributions are generated, one where for every wounded nucleon a contribution to the energy density with Gaussian shape and width $\sigma_0$ in both $x$ and $y$ is added, one where the same is done for every binary collision. These are then multiplied by $0.75$ and $0.25$, respectively, and added.

In the rapidity direction, we assume the energy density to be constant on a central plateau and fall like half-Gaussians at large $|\eta_s|$ as
described in \cite{Schenke:2010nt}:
\begin{equation}
  \varepsilon(\eta_s)\propto\exp\left[-\frac{(|\eta_s|-\eta_{\rm flat}/2)^2}{2 \sigma_{\eta}^2}\theta(|\eta_s|-\eta_{\rm flat}/2)\right]
\end{equation}

This procedure generates flux-tube like structures compatible with measured long-range rapidity correlations 
\cite{Jacobs:2005pk,Wang:2004kfa,Adams:2005ph}.
The absolute normalization is determined by demanding that the obtained total multiplicity distribution reproduces the experimental data.

As equation of state we employ the parametrization ``s95p-v1'' from \cite{Huovinen:2009yb}, obtained from interpolating between
lattice data and a hadron resonance gas.

\section{Freeze-out}\label{freeze}
We perform a Cooper-Frye freeze-out using
\begin{equation}\label{cf}
E\frac{dN}{d^3p}=\frac{dN}{dy p_T dp_T d\phi_p} = g_i \int_\Sigma f(u^\mu p_\mu) p^\mu d^3\Sigma_\mu\,,
\end{equation}
where $g_i$ is the degeneracy of particle species $i$, and $\Sigma$ the freeze-out hyper-surface.
In the ideal case the distribution function is given by
\begin{equation}
  f(u^\mu p_\mu) = f_0(u^\mu p_\mu) = \frac{1}{(2\pi)^3}\frac{1}{\exp((u^\mu p_\mu -\mu_i)/T_{\rm FO})\pm 1}\,,
\end{equation}
where $\mu_i$ is the chemical potential for particle species $i$ and $T_{\rm FO}$ is the freeze-out temperature.
In the finite viscosity case we include viscous corrections to the distribution function, $f=f_0+\delta f$, with
\begin{equation}\label{eq:delf}
 \delta f = f_0 (1\pm f_0) p^\alpha p^\beta \pi_{\alpha\beta} \frac{1}{2 (\epsilon+\mathcal{P}) T^2}\,,
\end{equation}
where $\pi$ is the viscous correction introduced in Eq. (\ref{tmunu}).
Note that the choice $\delta f \sim p^2$ is not unique \cite{Dusling:2009df} and a potential source of uncertainties particularly at the larger studied
 $p_T$.

The algorithm used to determine the freeze-out surface $\Sigma$ has been presented in \cite{Schenke:2010nt}. 
It is very efficient in determining the freeze-out surface of a system with fluctuating initial conditions. 

\section{Analysis and results}\label{results}
While in standard hydrodynamic simulations with averaged initial conditions all odd flow coefficients
vanish by definition, fluctuations generate all flow harmonics as response to the initial geometry.
We follow \cite{Petersen:2010cw}, where $v_3$ is computed in a similar way to the standard event plane analysis for elliptic flow,
and for each $v_n$ define an event plane through the angle
\begin{equation}
  \psi_n=\frac{1}{n}\arctan\frac{\langle \sin(n\phi)\rangle}{\langle \cos(n\phi)\rangle}\,.
\end{equation}
Note that here we do not weigh the average by $p_T$ as done in \cite{Petersen:2010cw,Schenke:2010rr} and \cite{Poskanzer:1998yz}.
This is because we focus on the low momentum part of $v_n$ in our analysis and hence 
want the event planes to be defined mainly using the low momentum part of the spectra. Differences between the different definitions are however small and lead to variations of $v_n$ on the order of one percent or less.

The flow coefficients can be computed using
\begin{equation}
  v_n=\langle \cos(n(\phi-\psi_n)) \rangle\,.
\end{equation}
When averaging over events we compute the root mean square $\sqrt{\langle v_2^2\rangle}$ because we compare to data obtained with the event-plane method 
(see \cite{Bhalerao:2006tp}).
First, we present results for particle spectra as functions of $p_T$ and $\eta_s$. Parameters were chosen in order to reproduce the experimental data for
the spectra when including all resonances up to $2\,{\rm GeV}$ (and some higher lying resonances to 
be consistent with what is included in the employed equation of state). The used parameters can be found in Table \ref{tab:parms}. 
Values for the maximal average energy density (in the center of the system) $\langle \varepsilon_{\rm max} \rangle$ are quoted for most central (0-5\%)
collisions. In addition, all parameter sets use $\eta_{\rm flat}=4.8$ and $\sigma_\eta=0.7$.

\begin{table}[htb]
  \centering
  \begin{tabular}{|c|c|c|c|c|c|}
    \hline
    $\eta/s$ & $\sigma_0 [{\rm fm}]$ & $\tau_0 [{\rm fm}/c]$ & $\langle \varepsilon_{\rm max} \rangle [{\rm GeV}/{\rm fm^3}]$
    & $T_{\rm FO} [{\rm MeV}]$    \\\hline
    0    & 0.4 & 0.4 & 65.7 & 150  \\
    0.08 & 0.4 & 0.4 & 57 & 150  \\
    0.16 & 0.4 & 0.4 & 50 & 150  \\
    0.08 & 0.2 & 0.4 & 57 & 155  \\
    0.08 & 0.8 & 0.4 & 57 & 145  \\
    \hline
  \end{tabular}
  \caption{Parameter sets. \label{tab:parms}}
\end{table}

\begin{figure}[htb]
   \begin{center}
     \includegraphics[width=8.5cm]{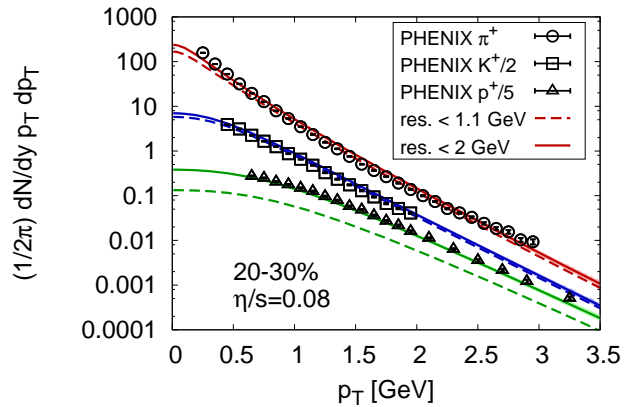}
      \caption{(Color online) Positive pion transverse momentum spectrum for 20-30\% central Au+Au collisions using $\eta/s=0.08$ including resonances up to $2\,{\rm GeV}$ (solid) and up to the $\phi$-meson (dashed) compared to data from PHENIX \cite{Adler:2003cb}. Results are averages over 10 single events.}
      \label{fig:pt-20-30}
   \end{center}
\end{figure}

\begin{figure}[htb]
   \begin{center}
     \includegraphics[width=8.5cm]{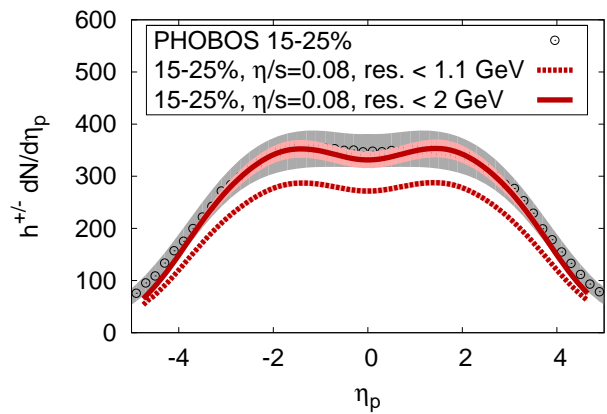}
      \caption{(Color online) Charged hadron spectrum for 15-25\% central Au+Au collisions including resonances up to $2\,{\rm GeV}$ 
        (solid, averaged over 10 events) and 
        up to the $\phi$-meson (dotted, averaged over 100 events) compared to data from PHOBOS \cite{Back:2002wb}.}
      \label{fig:eta}
   \end{center}
\end{figure}

Fig. \ref{fig:pt-20-30} shows the transverse momentum spectra of positive pions, kaons and protons compared to experimental data from PHENIX \cite{Adler:2003cb} in 20-30\% central events.
In Fig.\,\ref{fig:eta} we present a comparison of the computed charged particle spectrum for $\eta/s=0.08$ in 15-25\% central collisions as a function of pseudo-rapidity $\eta_p$ with experimental data from PHOBOS \cite{Back:2002wb}.

\begin{figure}[htb]
   \begin{center}
     \includegraphics[width=8.5cm]{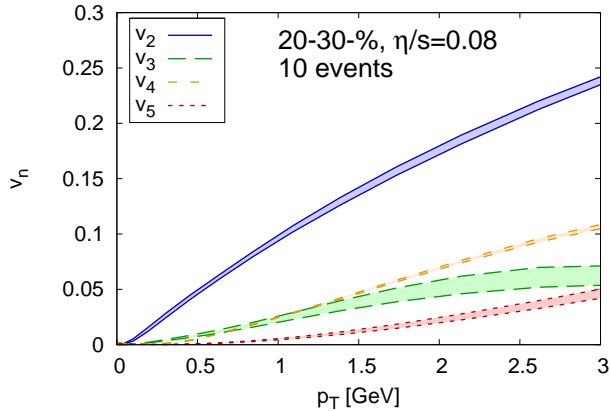}
      \caption{(Color online) Charged hadron $v_2$ to $v_5$ for $\eta/s=0.08$ as a function of transverse momentum $p_T$ averaged over 10 single events,
        including resonances up to the $\phi$-meson (upper end of each band) and all resonances up to $2\,{\rm GeV}$ (lower end of each band).}
      \label{fig:reso}
   \end{center}
\end{figure}

With the employed parameters we achieve very good agreement when including all resonance decays. In general, 
it is computationally too expensive to include resonances up to $2\,{\rm GeV}$ for all calculations. Hence, for most presented results 
we restrict ourselves to  
including resonances up to the $\phi$-meson only. This is a good approximation because
pions dominate the flow of all charged hadrons and it is mainly the $\rho$- and $\omega$- mesons that modify the pion distributions.
Fig. \ref{fig:reso} shows how the $v_n$ for charged hadrons are affected by including different numbers of resonances. 
Including more resonances reduces all $v_n$, however, the quantitative effect is small. 
The reduction is caused by the kinematics of resonance decays.
When including more resonances, at midrapidity one increases contributions from resonances at greater rapidities where 
the anisotropic flow is weaker.
The influence of higher lying resonances on $v_3$ appears to be larger than that on the other $v_n$. 

\begin{figure}[htb]
   \begin{center}
     \includegraphics[width=8.5cm]{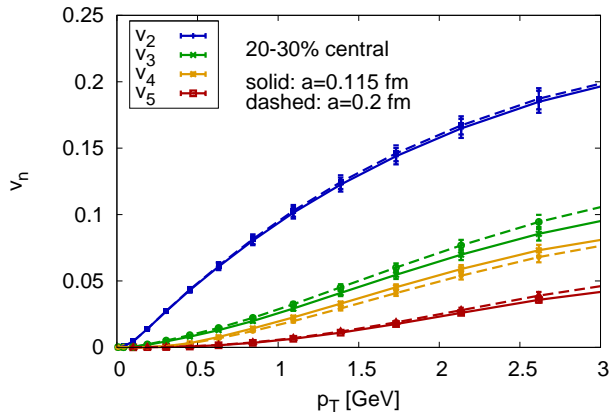}
      \caption{(Color online) Charged hadron $v_2$ to $v_5$ for $\eta/s=0.08$ and $\sigma_0=0.4\,{\rm fm}$ as a function of 
        transverse momentum $p_T$ averaged over 100 single events
        for lattice spacings $a=0.115\,{\rm fm}$ (solid lines) and $a=0.2\,{\rm fm}$ (dashed lines). }
      \label{fig:lat}
   \end{center}
\end{figure}

Next, we verify that our results are not plagued by large discretization errors. Higher flow harmonics are sensitive to fine structures in the system and 
for the case of ideal hydrodynamics with smooth initial conditions it was shown in \cite{Schenke:2010nt} that $v_4$ is very sensitive to the lattice spacing if it is not chosen small enough. Fig.\,\ref{fig:lat} shows $v_n(p_T)$ for two different lattice spacings, our standard value of $a=0.115\,{\rm fm}$ and a larger $a=0.2\,{\rm fm}$. Differences are within the statistical error bars from averaging over 100 events each. 

Because we are presenting the first (3+1)-dimensional relativistic viscous hydrodynamic simulation,
it is interesting to demonstrate the effect of shear viscosity on the longitudinal dynamics of the system, which
in a (1+1)-dimensional simulation was studied in \cite{Bozek:2007qt,Monnai:2011ju}.

\begin{figure}[th]
   \begin{center}
     \includegraphics[width=8.5cm]{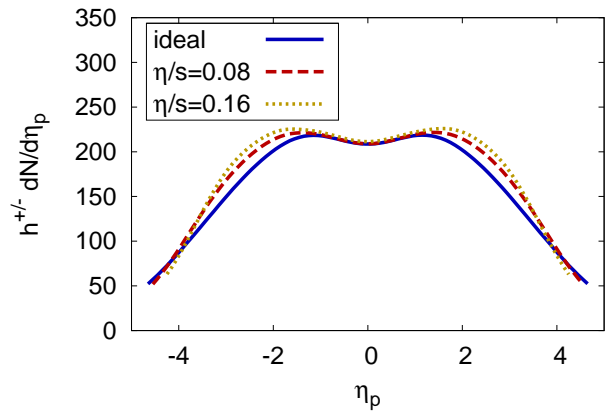}
      \caption{(Color online) Charged hadron spectrum for 20-30\% central Au+Au collisions for different values of $\eta/s$ including resonances 
        up to the $\phi$-meson.}
      \label{fig:Neta20-30}
   \end{center}
\end{figure}

\begin{figure}[th]
   \begin{center}
     \includegraphics[width=8.5cm]{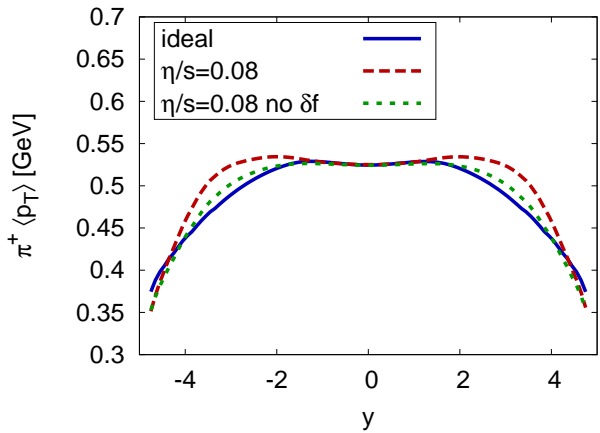}
      \caption{(Color online) Positive pion average $p_T$ as a function of rapidity $y$ for 20-30\% central Au+Au collisions 
        from ideal and viscous ($\eta/s=0.08$) including resonances up to the $\phi$-meson.}
      \label{fig:avPTeta20-30}
   \end{center}
\end{figure}

Fig.\,\ref{fig:Neta20-30} shows the modification of charged hadron pseudo-rapidity spectra caused by the inclusion of shear viscosity. 
The shape of the initial energy density distribution in the longitudinal direction is the same for all curves, which were each averaged over 200 events. 
The normalization was adjusted to yield the same multiplicity at midrapidity in all cases. In the range $2<|\eta_p|<4$ the 
pseudo-rapidity spectra are increased, for larger $\eta_p$ decreased by the effect of shear viscosity.
We checked that this effect is almost entirely due to the modified evolution when including shear viscosity. 
The viscous correction to the distribution functions $\delta f$ (\ref{eq:delf}) only causes minor modifications.
Additional information can be obtained by looking at the average transverse momentum $\langle p_T \rangle$ as a function of rapidity.
We show in Fig.\,\ref{fig:avPTeta20-30} that also $\langle p_T \rangle$ increases at intermediate rapidities and decreases at the largest $|y|$.
For this observable the effect of $\delta f$ is larger.

The modification in the viscous case is caused by the following effect: 
Faster longitudinal fluid cells drag slower neighbors by the viscous shear coupling. Naturally, the inclusion of both transverse
and longitudinal spatial dimensions in the simulation is needed to allow for such coupling.
Fast fluid elements are slowed down, slower ones sped up, decreasing the number of fluid cells with the largest rapidities, increasing the number at 
intermediate rapidities.
Further diffusion then distributes the momentum in all directions, explaining the increase of $\langle p_T\rangle$ at intermediate rapidities.

\begin{figure}[t]
  \begin{minipage}[b]{\linewidth}
    \centering
    \includegraphics[width=8.5cm]{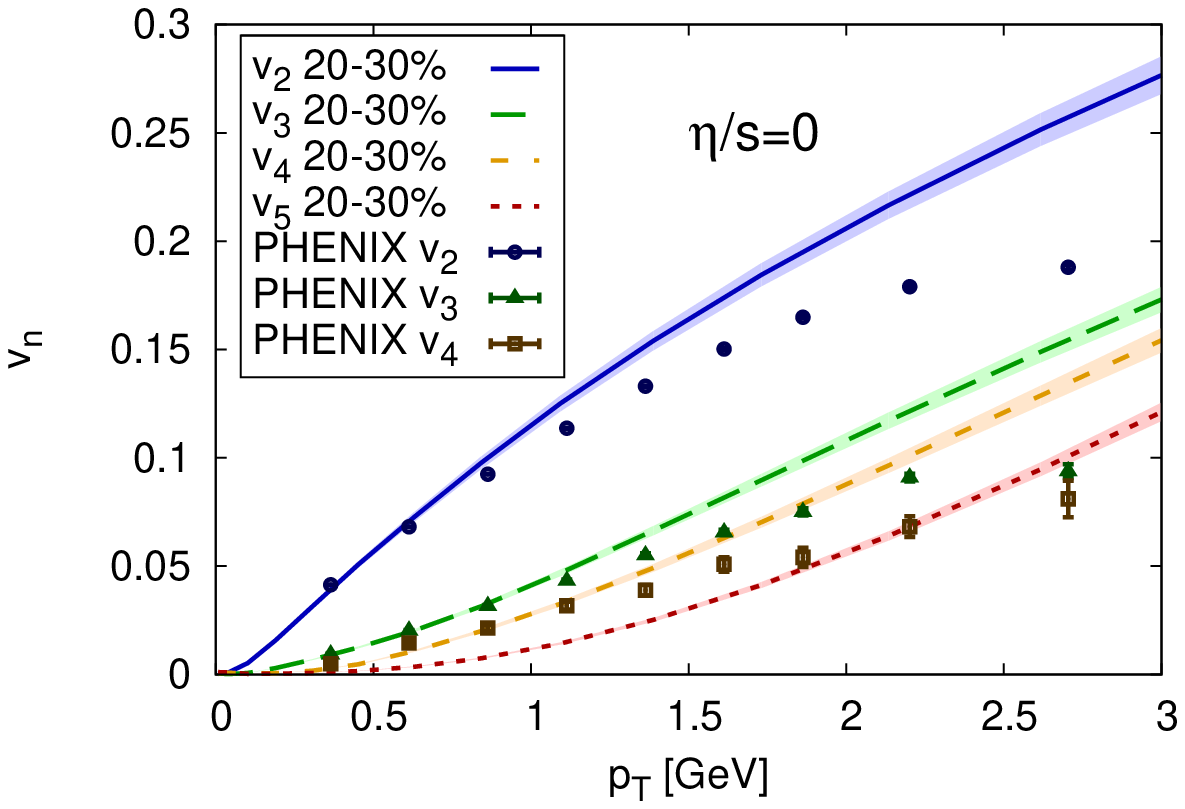}
  \end{minipage}\\
  \begin{minipage}[b]{\linewidth}
    \centering
    \includegraphics[width=8.5cm]{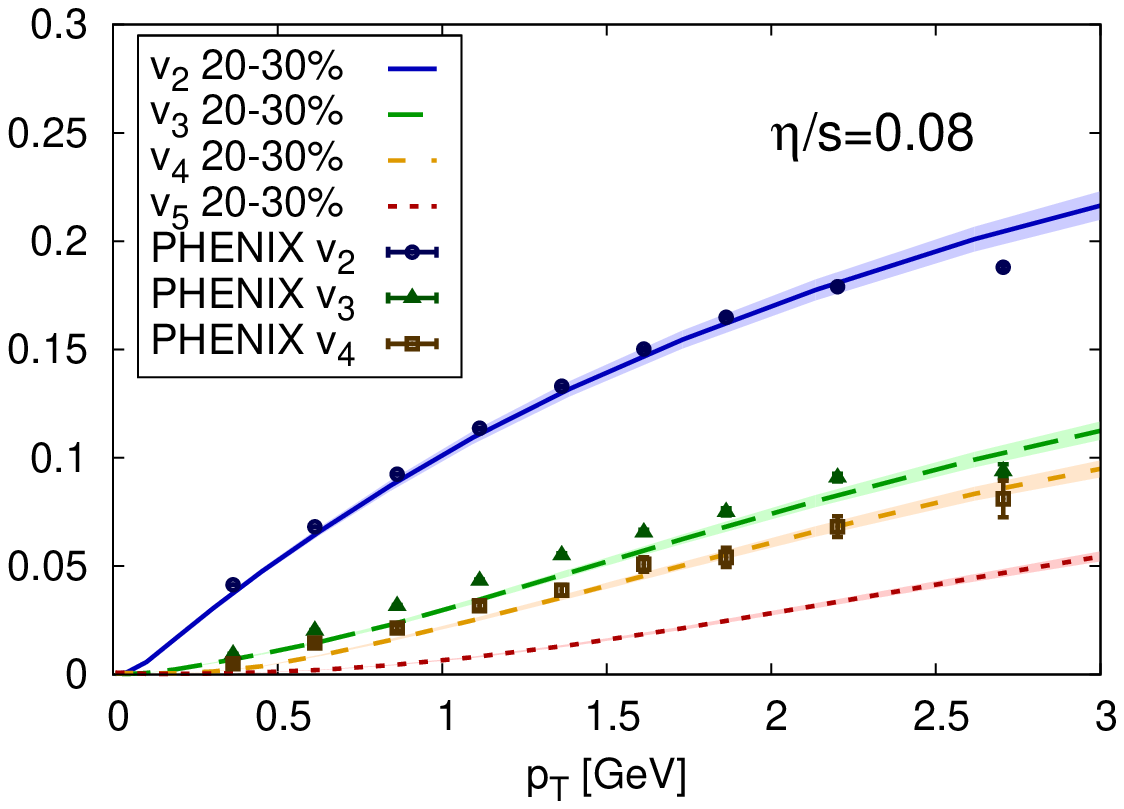}
  \end{minipage}\\
  \begin{minipage}[b]{\linewidth}
    \centering
    \includegraphics[width=8.5cm]{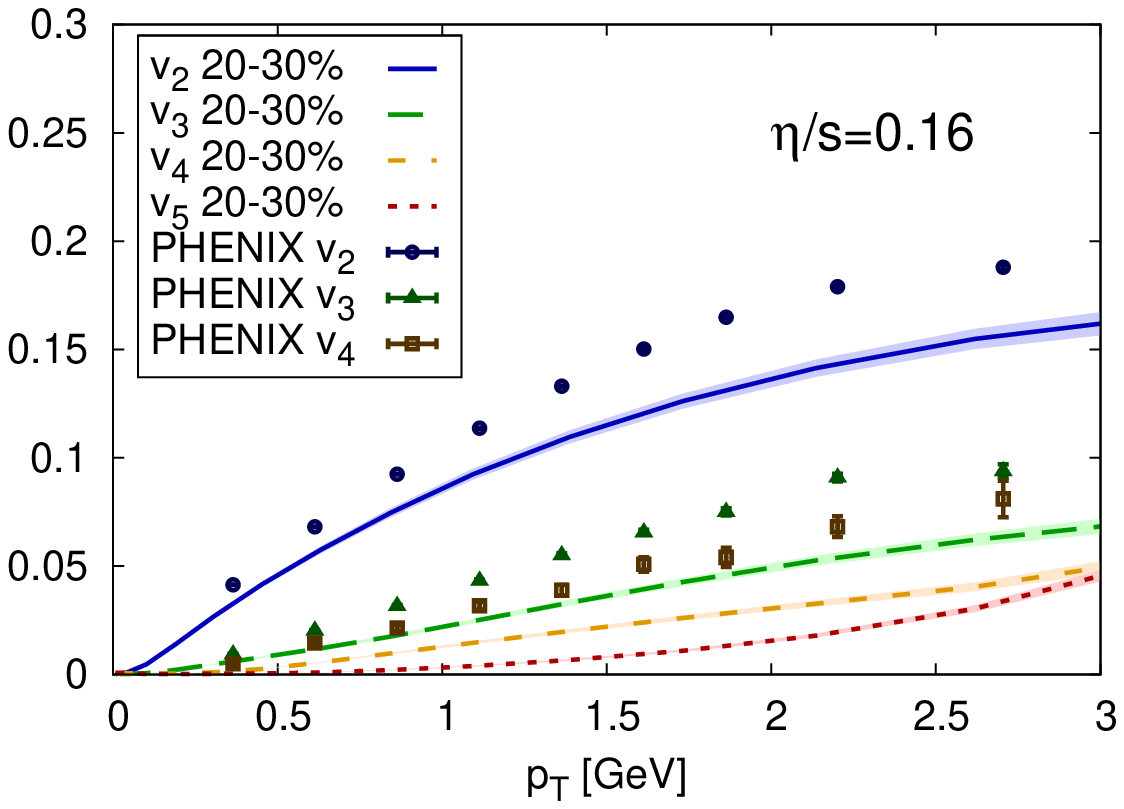}
  \end{minipage}
    \caption{(Color online) $p_T$-differential $v_2$ to $v_5$ from ideal hydrodynamics (left), viscous hydrodynamics with $\eta/s=0.08$ (middle),
      and $\eta/s=0.16$ (right). Results are averaged over 200 events each. Experimental data from PHENIX \cite{Adare:2011tg}. \label{fig:vn20-30}}
\end{figure}

\begin{figure}[htb]
  \begin{center}
    \includegraphics[width=8.5cm]{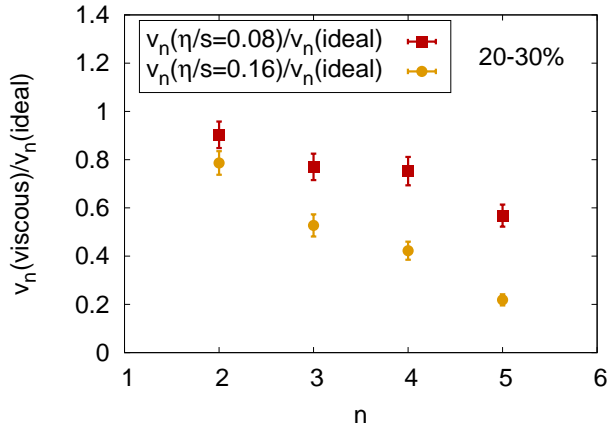}
    \caption{(Color online) Ratio of charged hadron flow har\-mo\-nics in viscous simulations to the result from ideal hydrodynamics.
      Results are averages over 200 single events each.}
    \label{fig:vntot20-30ratio}
  \end{center}
 \end{figure}

In Fig.\,\ref{fig:vn20-30} we show the dependence of $v_n(p_T)$ on the shear viscosity of the system. 
Results are averaged over 200 single events each. 
For $v_2$ to $v_4$ we compare to experimental data from the PHENIX collaboration obtained using the event plane method \cite{Adare:2011tg}.
One can clearly see that the dependence of $v_n(p_T)$ on $\eta/s$ increases with increasing $n$. To make this point more quantitative, 
we present the ratio of the $p_T$-integrated $v_n$ from viscous calculations to $v_n$ from ideal calculations as a function of $n$ in Fig.\,\ref{fig:vntot20-30ratio}. 
While $v_2$ is suppressed by $\sim 20\%$ when using $\eta/s=0.16$, $v_5$ is suppressed by $\sim 80\%$. Higher harmonics are substantially more affected
by the system's shear viscosity than $v_2$ and hence are a much more sensitive probe of $\eta/s$.
This behavior is expected because diffusive processes smear out finer structures corresponding to higher $n$ more efficiently than larger scale structures,
and has been pointed out previously in \cite{Alver:2010dn}.


So far all results were obtained using initial conditions with a Gaussian width $\sigma_0=0.4\,{\rm fm}$. We now study the effect of 
the initial state granularity on the flow harmonics by varying $\sigma_0$. Decreasing $\sigma_0$ causes finer structures to appear and hence 
strengthens the effect of hot spots. This results in a hardening of the spectra as previously demonstrated in \cite{Holopainen:2010gz}.
Because we want to compare to experimental data, we readjust 
the slopes to match the experimental $p_T$-spectra by modifying the freeze-out temperature (see Table \ref{tab:parms}).

\begin{figure}[htb]
  \begin{minipage}[b]{\linewidth}
    \centering
     \includegraphics[width=8.5cm]{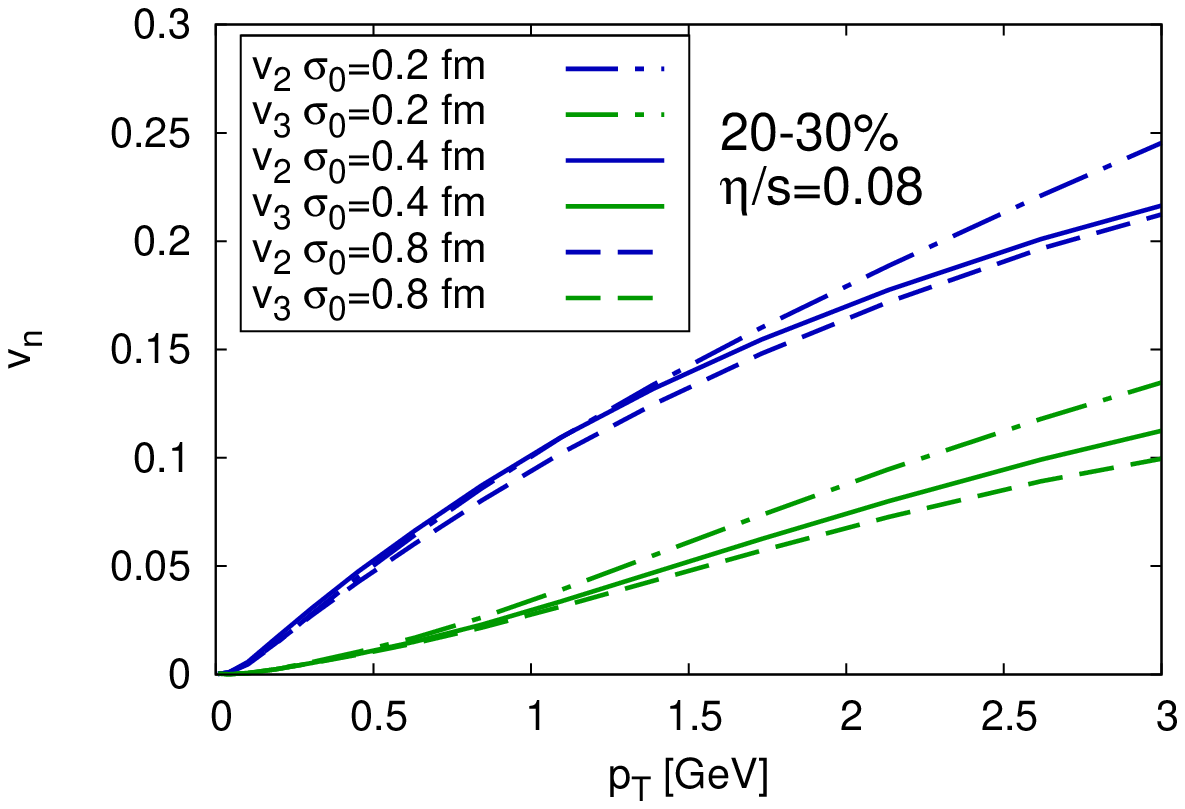}
      \label{fig:v2-v3-20-30sigma}
  \end{minipage}\\
  \begin{minipage}[b]{\linewidth}
    \centering
     \includegraphics[width=8.5cm]{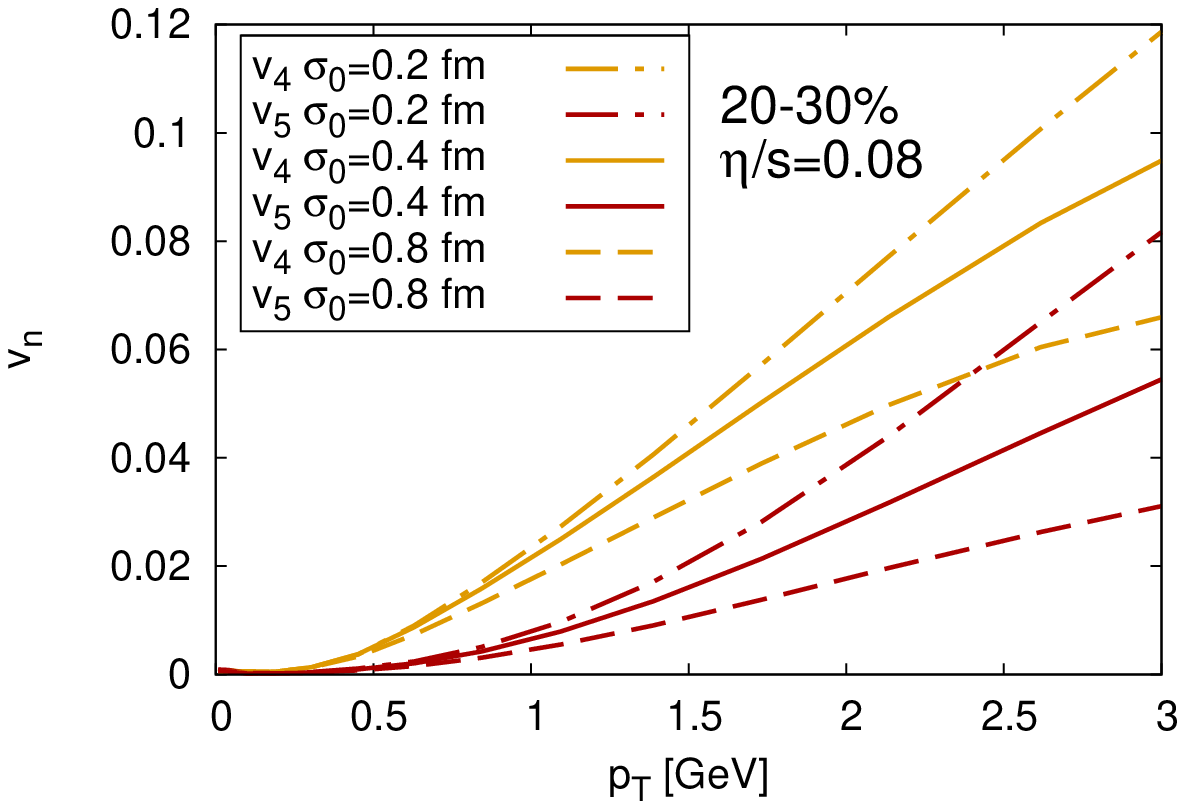}
   \end{minipage}
    \caption{(Color online) Differential $v_2$ and $v_3$ (upper panel) and $v_4$ and $v_5$ (lower panel) 
        in 20-30\% central collisions using $\eta/s=0.08$ and varying $\sigma_0$.
        Results are averages over 100 single events each (200 events for $\sigma_0=0.4\,{\rm fm}$).     \label{fig:v4-v5-20-30sigma}}
 \end{figure}

\begin{figure*}[htb]
  \begin{minipage}[t]{0.45\linewidth}
    \centering
    \includegraphics[width=8.5cm]{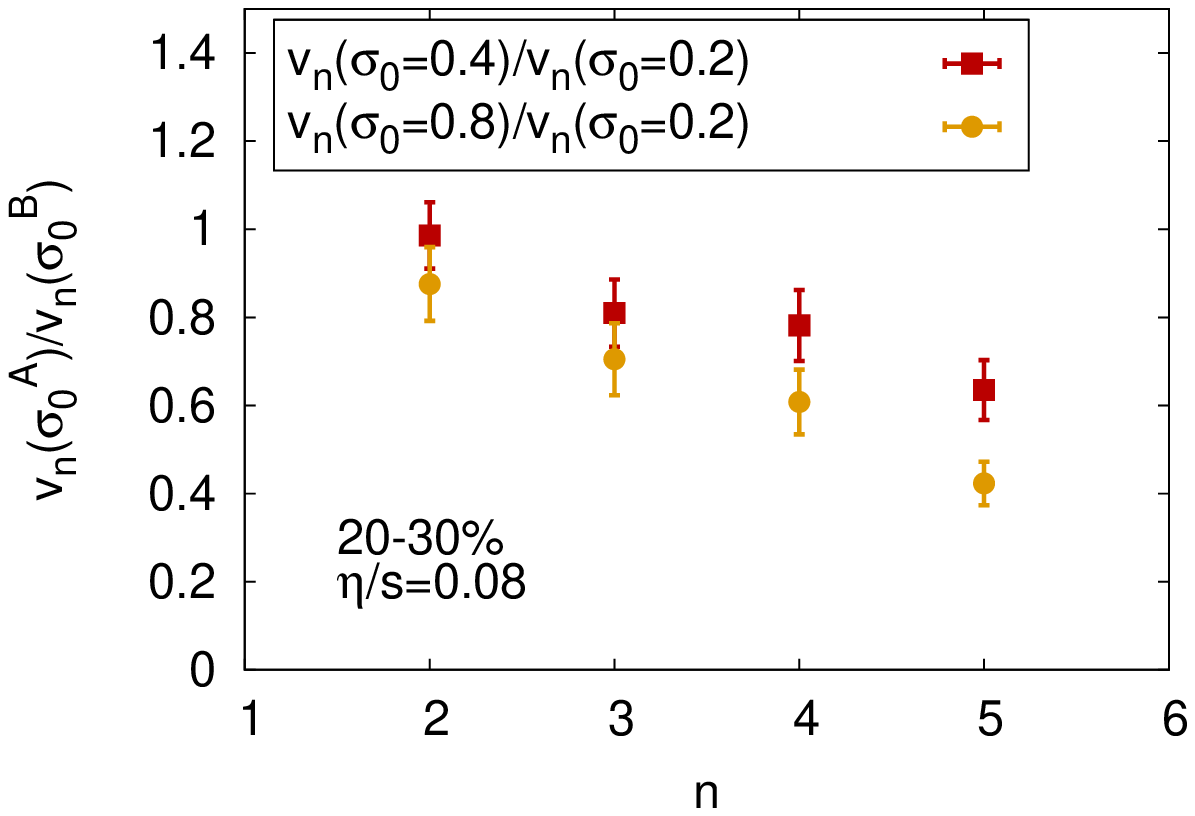}
    \caption{(Color online) Ratio of $v_n$ with initial granularity characterized by the Gaussian width $\sigma_0=0.8\,{\rm fm}$
      to the case with $\sigma_0=0.4\,{\rm fm}$ and $\sigma_0=0.8\,{\rm fm}$, respectively.
      Results are for 20-30\% central collisions using $\eta/s=0.08$. Averages are over 100 single events each.}
    \label{fig:vntot20-30sigmaratio}
  \end{minipage}
  \hspace{0.8cm}
  \begin{minipage}[t]{0.45\linewidth}
    \centering
     \includegraphics[width=8.5cm]{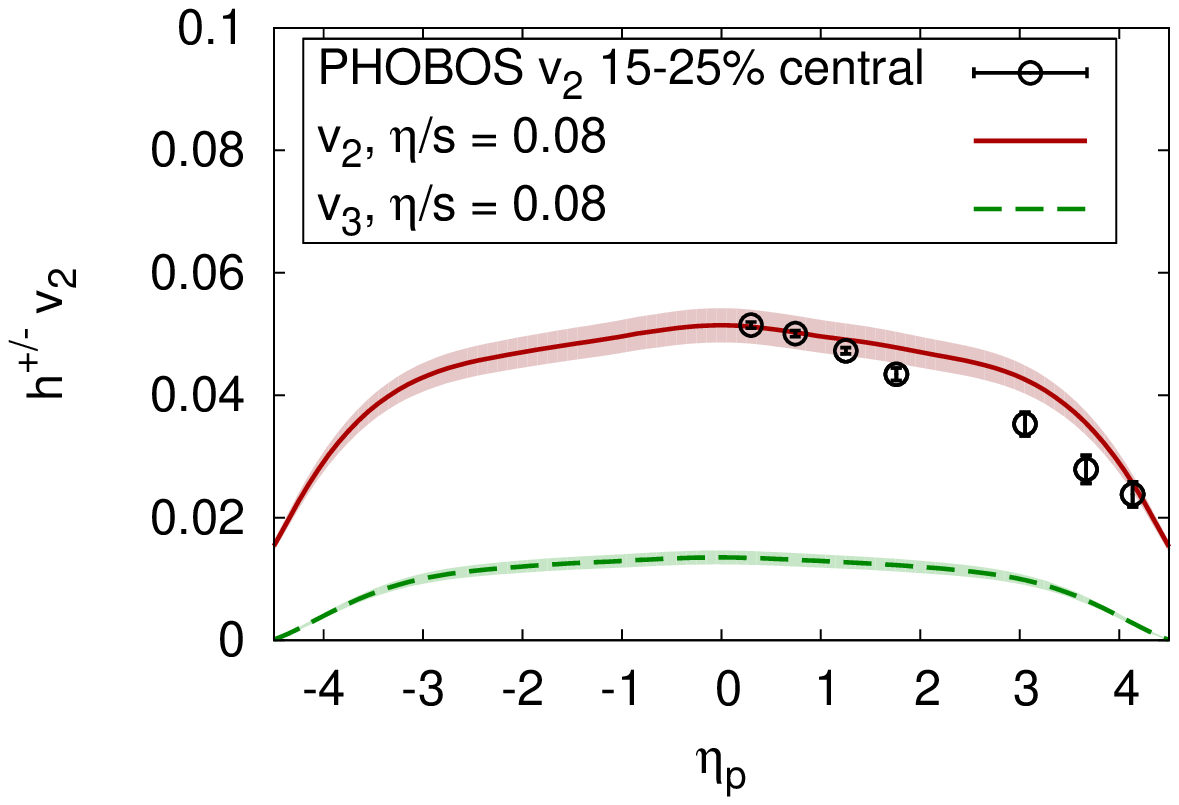}
      \caption{(Color online) $v_2$ and $v_3$ as functions of pseudo-rapidity $\eta_p$ compared to data from PHOBOS \cite{Back:2004mh}.
        Averages are over 100 single events each.}
      \label{fig:v2eta-h}
  \end{minipage}
\end{figure*}

Fig.\,\ref{fig:v4-v5-20-30sigma} shows the dependence of $v_n(p_T)$ on the value of $\sigma_0$, which we vary from $0.2\,{\rm fm}$ to $0.8\,{\rm fm}$. 
While $v_2$ is almost independent of $\sigma_0$, higher flow harmonics show a very strong dependence.
In Fig.\,\ref{fig:vntot20-30sigmaratio} we present the dependence of the $p_T$-integrated $v_n$ on the initial state granularity characterized by $\sigma_0$.

Higher flow harmonics turn out to be a more sensitive probe of initial state granularity than $v_2$.
While we are not yet attempting an exact extraction of $\eta/s$ using higher flow harmonics, our results give a first quantitative 
overview of the effects of both the initial state granularity and $\eta/s$ on all higher flow harmonics up to $v_5$. 
Comparing Figs.\,\ref{fig:vn20-30} and \ref{fig:v4-v5-20-30sigma}, we see that $v_4(p_T)$ obtained from simulations using $\eta/s=0.16$ is about a factor of 2 below the experimental result, and that decreasing $\sigma_0$ by a factor of two does not increase it nearly as much.
Note that $\sigma_0=0.2\,{\rm fm}$ is already a very small value given that we assign this width to a wounded nucleon. 
It is hence unlikely that a higher initial state granularity will be able to compensate for the large effect of the shear viscosity. 
Similar arguments hold for $v_3(p_T)$. 

A detailed systematic analysis of different mo\-dels for the initial state with a sophisticated description of fluctuations is needed to make more precise
statements on the value of $\eta/s$. It is however clear from the present analysis that the utilization of higher flow harmonics can constrain
models for the initial state and values of transport coefficients of the quark-gluon plasma significantly. 
The analysis of only elliptic flow is not sufficient for this task, because it depends too weakly on both the initial state granularity and $\eta/s$.



\begin{figure*}[htb]
  \begin{minipage}[b]{0.5\linewidth}
    \centering
    \includegraphics[width=8.5cm]{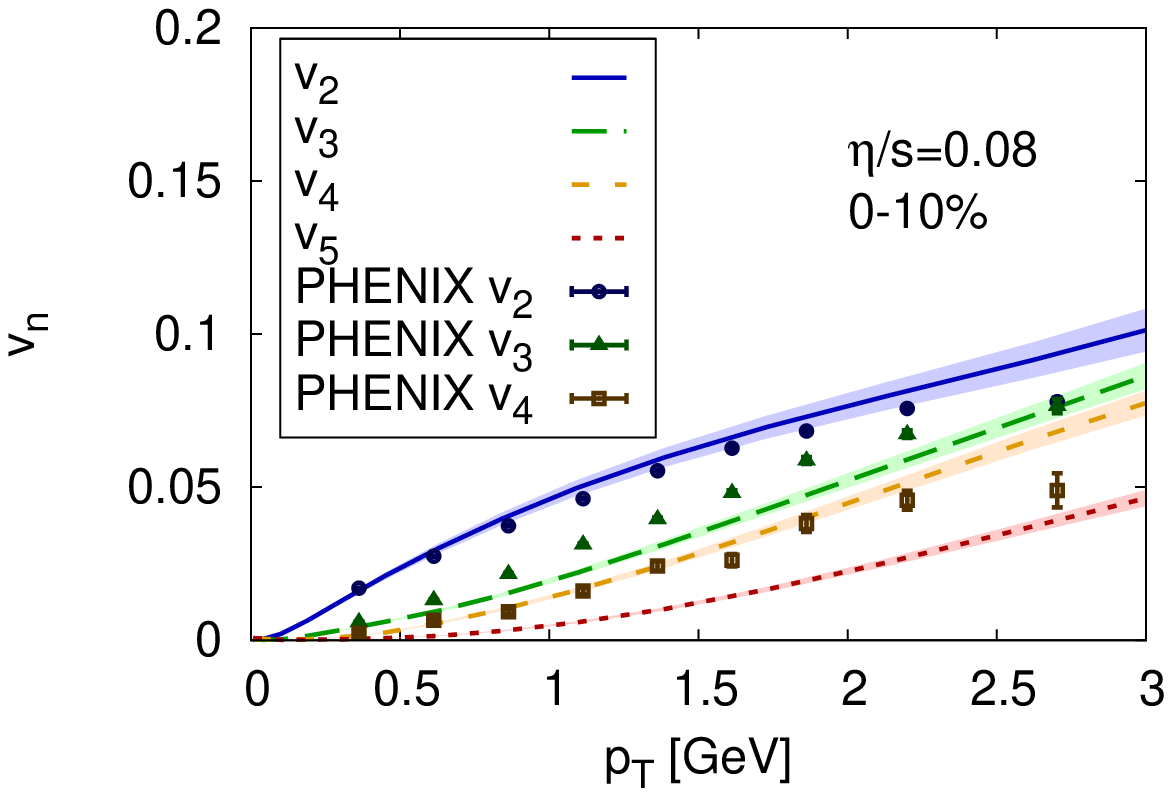}
  \end{minipage}
  \hspace{-0.5cm}
  \begin{minipage}[b]{0.5\linewidth}
    \centering
    \includegraphics[width=8.5cm]{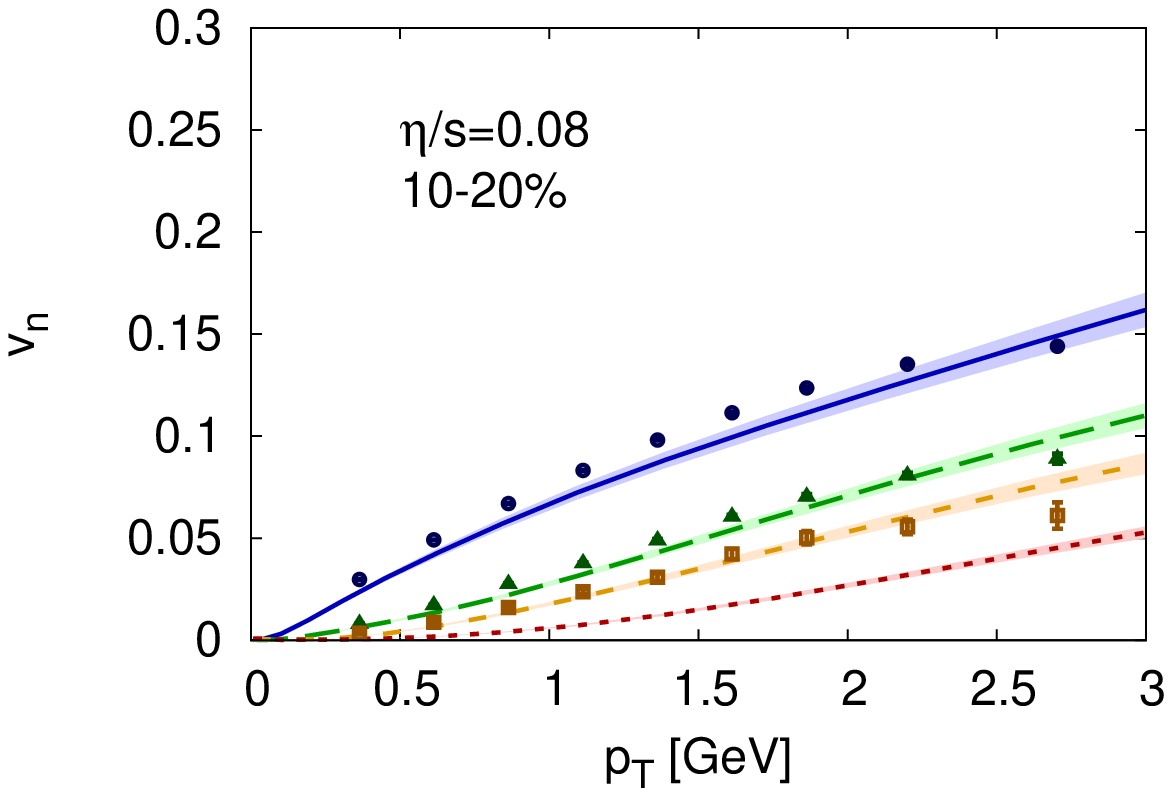}
  \end{minipage}\\
  \hspace{-0.5cm}
  \begin{minipage}[b]{0.5\linewidth}
    \centering
    \includegraphics[width=8.5cm]{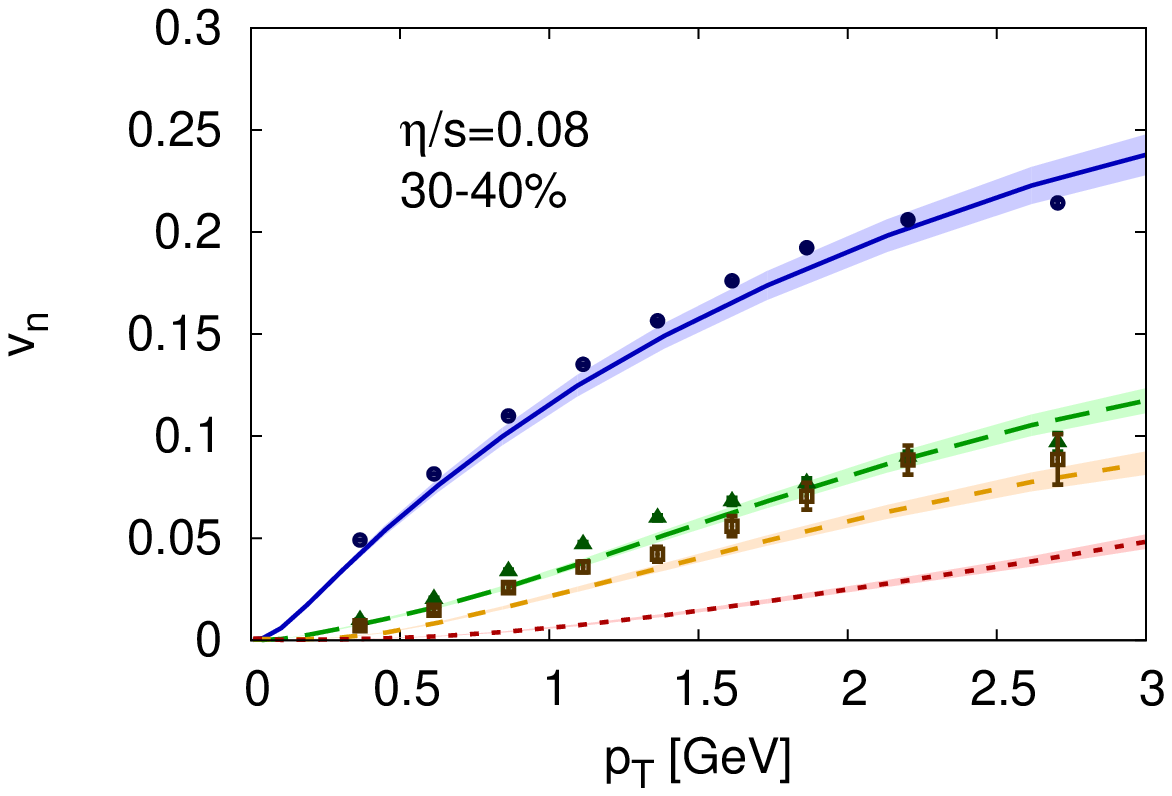}
  \end{minipage}
  \hspace{-0.5cm}
  \begin{minipage}[b]{0.5\linewidth}
    \centering
    \includegraphics[width=8.5cm]{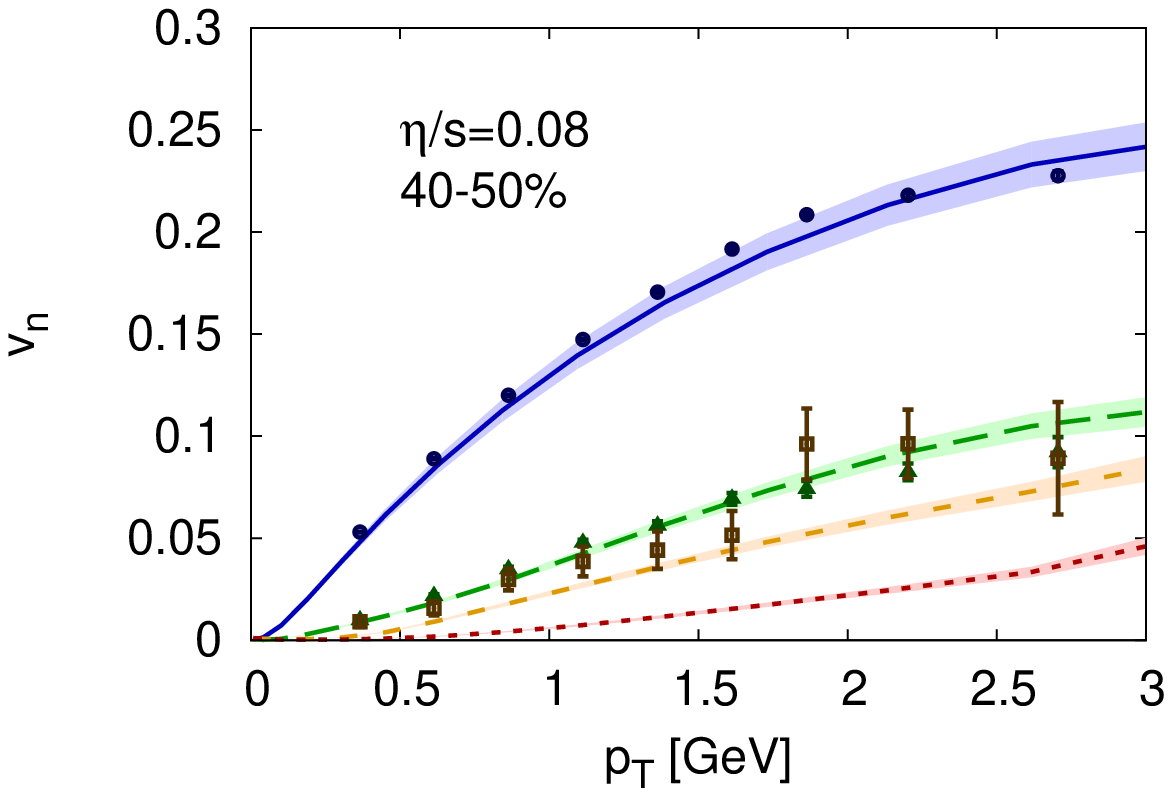}
  \end{minipage}
    \caption{(Color online) $p_T$-differential $v_2$ to $v_5$ from viscous hydrodynamics with $\eta/s=0.08$ for centralities 0-10\% (upper left),
      10-20\% (upper right), 30-40\% (lower left), and 40-50\% (lower right). See Fig.\,\ref{fig:vn20-30} for 20-30\% central collisions.
      Results are averaged over 100 events each. Experimental data from PHENIX \cite{Adare:2011tg}. \label{fig:vncent}}
\end{figure*}

We present $v_2$ and $v_3$ as a function of pseudo-rapidity in Fig.\,\ref{fig:v2eta-h}. The $v_2(\eta_p)$ result from the simulation is 
flatter than the experimental data out to $\eta_p\approx 3$ and then falls off more steeply. A modified shape of the initial energy density distribution
in the $\eta_s$-direction, the inclusion of finite baryon number, and 
inclusion of a rapidity dependence of the fluctuations will most likely improve the agreement.

In Fig.\,\ref{fig:vncent} we show results of $v_n(p_T)$ for different centralities using $\eta/s=0.08$. 
Overall, all flow harmonics are reasonably well reproduced. Deviations from the experimental data, especially of $v_3(p_T)$ in the most central 
collisions indicate that our rather simplistic description of the initial state and its fluctuations is insufficient.
Improvements can be made by a systematic study with alternative models for the fluctuating 
initial state based on e.g. the color-glass-condensate effective theory (along the lines of \cite{Drescher:2007ax}).

\begin{figure}[htb]
   \begin{center}
     \includegraphics[width=8.5cm]{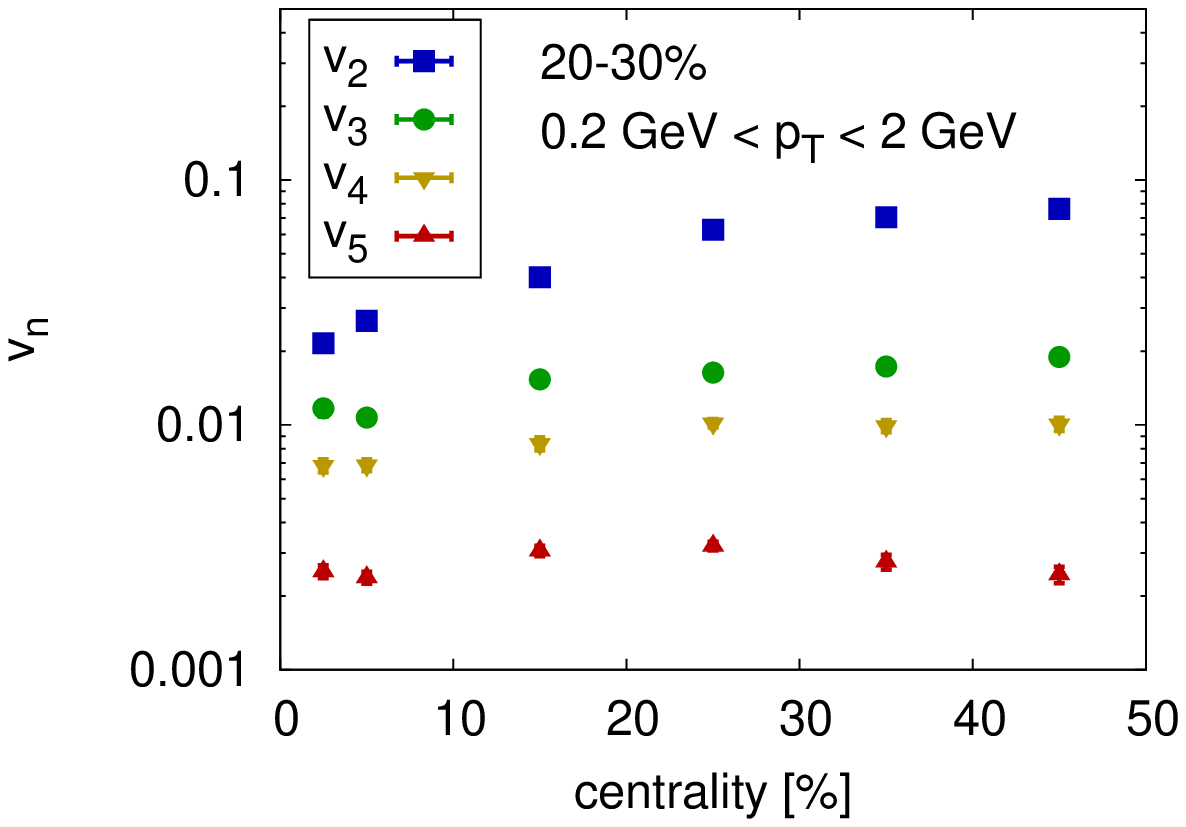}
      \caption{(Color online) $v_2$ to $v_5$ as functions of centrality for $\eta/s=0.08$.
        Averages are over 100 single events each.}
      \label{fig:vntotcent}
   \end{center}
\end{figure}

Finally, the higher flow harmonics integrated over a transverse momentum range $0.2\,{\rm GeV} < p_T < 2\,{\rm GeV}$ 
are shown in Fig.\,\ref{fig:vntotcent} as a function of centrality. $v_2$ has the strongest dependence 
on the centrality because it is driven to a large part by the overall geometry. 
The odd harmonics are entirely due to fluctuations as we have discussed earlier, and hence do not show a strong dependence on the centrality of the collision.

\section{Summary and Conclusions}\label{summary}
We have demonstrated that the analysis of higher flow harmonics within (3+1)-dimensional 
event-by-event viscous hydrodynamics has the potential to determine
transport coefficients of the QGP such as $\eta/s$ much more precisely than the analysis of elliptic flow alone.
We presented in detail the framework of (3+1)-dimensional viscous relativistic hydrodynamics and introduced the concept of event-by-event 
simulations, which enable us to study quantities that are strongly influenced or even entirely due to fluctuations such as odd flow harmonics.
Parameters of the hydrodynamic simulation were fixed to reproduce particle spectra both as a function of transverse momentum $p_T$ and pseudo-rapidity $\eta_p$. 
The studied flow harmonics $v_2$ to $v_5$ were found to depend increasingly strongly on the value of $\eta/s$ and also on the initial state granularity. 
This work does not attempt
an exact extraction of $\eta/s$ of the QGP but our quantitative results hint at a value of $\eta/s$ not larger than $2/4\pi$. The reason is the strong
suppression of $v_3$ to $v_5$ by the shear viscosity. A higher granularity of the initial state counteracts this effect, but our results indicate that 
this increase is not large enough to account for $\eta/s \geq 2/4\pi$.
We will report on a detailed analysis of higher flow harmonics at LHC energies and a comparison to the experimental data in a subsequent work.

\subsection*{Acknowledgments}
BPS thanks Roy Lacey and Raju Venugopalan for very helpful discussions.
This work was supported in part by the Natural Sciences and Engineering Research Council of Canada.
BPS is supported by the US Department of Energy under DOE Contract No.DE-AC02-98CH10886 and by a Laboratory Directed Research and
Development Grant from Brookhaven Science Associates. 
We gratefully acknowledge computer time on the Guillimin cluster at the CLUMEQ HPC centre, a part of Compute Canada HPC facilities.

\bibliography{hydro}

\end{document}